# Nano Josephson Superconducting Tunnel Junctions in Y-Ba-Cu-O Direct-Patterned with a Focused Helium Ion Beam

**Authors:** Shane A. Cybart,[1,2,*] E. Y. Cho,[1] T. J. Wong,[1] Björn H. Wehlin,[1] Meng K. Ma,[1] Chuong Huynh,[3] and R. C. Dynes[1,2]

**Affiliations:**

[1] Oxide Nano Electronics Laboratory, Department of Physics, University of California San Diego, La Jolla, California 92093, USA.

[2] Materials Sciences Division, Lawrence Berkeley National Laboratory, Berkeley, California 94720, USA.

[3] Carl Zeiss Microscopy, LLC., One Corporation Way, Peabody, Massachusetts 01960 USA.

*scybart@ucsd.edu

Since the discovery of the unconventional copper-oxide high-transition-temperature superconductors (HTS), researchers have explored many methods to fabricate superconducting tunnel junctions from these materials for both superconducting electronics operating at the practical temperature of liquid nitrogen (77 K) and for fundamental measurements essential for testing and guiding theories of these remarkable superconductors. The difficulty is that the traditional estimate of the superconducting coherence length is very short and anisotropic in these materials, typically ~2 nm in the *a-b* plane and ~0.2 nm along the *c*-axis. The coherence volume encloses very few superconducting pairs, so even the presence of small scale inhomogeneities can locally disrupt superconductivity unlike in conventional superconductors. Therefore the electrical properties of Josephson junctions are sensitive to chemical variations and structural defects on atomic length scales[1], thus to make multiple uniform HTS junctions, control at the atomic level is required. In this letter, we demonstrate very high-quality all-HTS Josephson superconducting tunnel junctions (both Josephson and quasiparticle tunneling) created by using a 500 pm diameter focused beam of helium ions to direct-write tunnel barriers into $YBa_2Cu_3O_{7-\delta}$ (YBCO) thin films. With this method we demonstrate the ability to control the barrier properties continuously from conducting to insulating by varying the irradiation dose. This technique provides a reliable and reproducible pathway for the scaling up of quantum mechanical circuits operating at practical temperatures (~77 K) as well as an avenue to conduct superconducting tunneling studies in HTS for basic science.

There are a variety of scientific and technological reasons why it would be significant to be able to fabricate large numbers of reproducible, high-quality Josephson junctions from HTS. Large-scale circuits with millions of Josephson junctions on a chip could open new applications in high-performance computing, high-frequency sensors, and magnetometery. Superconducting tunnel junctions also enable the spectroscopic study of HTS materials allowing access to direct symmetry and excitation spectroscopies.

Historically, typical HTS Josephson junctions have exhibited superconductor-normal metal-superconductor (SNS) properties, and superconductor-insulator-superconductor (SIS) tunneling in an all-HTS junction is rare but has been observed in mechanical break junctions[2], some grain boundary junctions[3], and multi-layer *c*-axis sandwich junctions[4-6]. These types of studies have provided a great deal of insight to theorists working on HTS, however, reproducibly fabricated tunnel junctions where the direction of transport can be controlled has eluded researchers due to the lack of a method to fabricate barriers with well-defined interfaces in the *a-b* plane small enough for tunneling to occur. To achieve this we have utilized a 500-pm diameter focused helium ion beam[7] to fabricate via irradiation damage a very narrow (~nm) tunnel barrier between two superconducting YBCO electrodes. The key to this method is that YBCO is very sensitive to disorder from irradiation and undergoes a transition from superconductor to an insulator as irradiation dose is increased[8]. Therefore varying the dose allows for precise control of junction properties.

Both electron beam irradiation[9,10] and masked ion implantation[11-13] have been used in the past as methods to fabricate Josephson SNS junctions but the dimensions were too large for SIS tunneling. These types of junctions exhibit small characteristic voltages $V_C = I_C R_N$ (a figure of merit for Josephson junctions equal to the product of the critical current and the normal state resistance) that precludes their use in most applications. $I_C R_N$ is small for these earlier junctions, because the irradiated weak links are much larger in the transport direction than the superconducting coherence length or a tunneling length. The typical width of the trenches used in the high-aspect-ratio masks used for defining these barriers is ~25 nm[14], and lateral straggle of defects from the implantation process broadens out the barrier so that the actual width of the irradiated region can approach 100 nm[15]. Josephson currents can only propagate through such large regions via the superconducting proximity effect, which is a phenomenon where non-superconducting materials in close electrical proximity with a superconductor become superconducting themselves. In the case of ion irradiated weak links the coupled materials are the same, but the irradiated region has a reduced transition temperature $T'_C$. If the irradiated region is narrow (< 100 nm) it will sustain a Josephson current above $T'_C$ but the pair potential $\Delta$ is significantly reduced from that of the electrodes which results in smaller values of $I_C R_N$. To obtain higher $I_C R_N$ the barrier needs to be about an order of magnitude smaller and lateral straggle of defects must be minimized. We have achieved this by reducing the thickness of the film in the device region such that the ion beam propagates through the film and embeds deep in the substrate leaving a narrow damage track. (Extended Data Fig. 1).

Another historical drawback to ion irradiated Josephson junctions is the presence of a large non-Josephson "excess current" at zero voltage that does not exhibit either the DC or AC Josephson effects. The physical origin of the excess current is understood in the framework of the Blonder, Tinkham, and Klapwijk model (BTK) for microscopic electrical transport at an interface between a superconductor and a normal material[16]. The power of this model is that it can describe current-voltage characteristics for barriers ranging from a strong barrier, such as an insulator in a tunnel junction, to a weak barrier like a normal metal, using a single parameter related to barrier strength. In the case of a strong barrier (a tunnel barrier) the only transport mechanism for cooper pairs is direct Josephson tunneling whereas in the case of weaker barriers both tunneling and Andreev reflection occur. Therefore to maximize the Josephson current and reduce excess current a strong barrier is required but it must also be confined to less than a few nanometers wide in order for tunneling to occur as the tunneling probability depends exponentially on the insulator thickness. This dimension is too challenging for most nanofabrication techniques such as electron beam lithography or gallium focused ion beams but as we show in this paper, it is straight-forward, for state-of-the-art focused helium ion beams from gas field-ion sources[7].

To fabricate focused helium ion beam Josephson junctions, large circuit features for electrical contacts, and 4-µm wide strips of YBCO were patterned with conventional photolithography in a YBCO thin film that had an *in situ* deposited Au contact layer on top (Fig. 1a). The starting YBCO film thickness was 150 nm, but the Au was removed and the YBCO was etched to a thickness of ~30 nm in the area intended for junctions (Fig. 1b) Samples were then loaded into a Zeiss Orion helium ion microscope and the 30 kV helium beam was scanned across the 4-µm wide superconducting bridges to form the tunnel barriers (Fig. 1c). Over 30 test samples were written with ion fluence ranging between $10^{14}$ and $10^{18}$ $He^+/cm^2$. At the lower values very little Josephson current was observed and SNS Andreev reflections dominated transport (Extended Data Fig. 2). In contrast, at the higher doses the devices exhibited strong localization characteristics with insulating behavior in the barrier (Extended Data Fig. 3). In between these two extremes we were able to determine doses that could create very high-quality Josephson junctions and continuously transition, from SNS junctions, well-described by the resistively shunted junction model (RSJ)[17,18], to SIS Josephson tunnel junctions.

The current-voltage characteristics (*I-V*) are shown in Fig. 2a for a typical SNS Josephson junction measured at several temperatures. The resistance is ~1 Ω and is roughly 10 times larger than typical ion irradiated weak links for the same bridge width[19]. Figure 2b shows the temperature dependence of $I_C$ and $R_N$. The decreasing resistance with decreasing temperature indicates the barrier is a conductor despite the dose, $2 \times 10^{16}$ $He^+/cm^2$, used to create this junction is about 10 times higher than that required to drive YBCO insulating when delivered to a large area[20]. We believe that the difference is due to the overlap of defects from lateral straggle when doses are delivered to a two-dimensional area as opposed to a one-dimensional line. The critical current substantially increases with decreasing temperature which is typical for SNS junctions because the barrier is becoming a stronger superconductor at lower temperature[16]. The temperature range for RSJ characteristics (~30 K) is much larger than that for proximity effect ion irradiated junctions (~3 K)[19] suggesting that the strength of the barrier is much higher and more efficient at suppressing the Andreev reflection (excess current) transport mechanism. The Fraunhofer diffraction pattern of the supercurrent in applied magnetic field (Fig. 2c) clearly demonstrates the dc Josephson effect, however the pattern is skewed due to self-field effects from the very high current density of 100 kA/cm².

In stark contrast to the SNS junction, Fig. 2d shows *I-V* for several temperatures of a YBCO, SIS junction fabricated by irradiation simply with a higher dose of $6\times10^{16}$ $He^+/cm^2$. Unlike the junction shown in Fig. 2a, the resistance of this junction increases as temperature is decreased indicating that the barrier is an insulator (Fig. 2e). Furthermore, unlike the SNS junction, the critical current more weakly depends on temperature as expected for a strong insulating barrier (Fig. 2e). It slightly increases as the thermal noise is reduced and fluctuations are suppressed. Measurements of the dynamic resistance (d*V*/d*I*) reveal 5 orders of AC Josephson resonant cavity modes (Fiske steps) generated at finite voltage due to the Earth's magnetic field with period of 35 µV (Extended Data Fig. 4). This implies that there is a well-defined dielectric barrier and the ac Josephson effect couples to the cavity defined by that barrier. The Fraunhofer diffraction pattern for this junction is shown in (Fig. 2f). The amplitudes of the side lobes fall off much slower than the patterns from ideal sandwich junctions because the current is concentrated closer to the edges. This effect was predicted by Humphreys and Edwards[21] and later by Clem[22] in their work describing the magnetic field characteristics of planar Josephson junctions.

At higher current bias (Fig. 2g) the SIS nature of the *I-V* is more apparent and conductance peaks are visible at $V = \pm32$ mV reflecting superconducting energy gap behavior.

We measure the differential resistance using low frequency techniques and d$I$/d$V$ is plotted in Fig. 2h. We define the conductance peaks to represent a superconducting energy gap, 2Δ and plot it as a function of temperature (Fig. 2i). This data fits surprisingly well to the BCS gap temperature dependence with only the two parameters 2Δ($T$ = 0 K) = 33 meV and $T_C$ = 77.8 K. The conductance decrease above 2Δ is a common occurrence in HTS tunnel junctions unlike those made from conventional superconductors. We speculate that in our devices this is due to an additional voltage caused by flux flow in the electrodes from the extremely high current density ~ 100MA/cm$^2$. To support this notion we measured $I$-$V$ above the critical current of the electrodes for many temperatures and show that this negative curvature increases rapidly up to the critical current of the electrodes (Extended Data Fig. 5).

We believe this new advance will have a significant and far reaching impact for applications of superconducting electronics covering a wide spectrum, ranging from highly sensitive magnetometers for biomagnetic measurements of the human body, to large scale arrays for wideband satellite communications[23]. For basic science, it will contribute to unraveling the mysteries of unconventional superconductors and could play a major role in new technologies such as quantum information science. Furthermore, the method of direct patterning thin films is not just limited to cuprate superconductors. This technique will be applicable to other superconductors like magnesium diboride[24] or materials that are sensitive to disorder, such as multiferroics, graphene, manganites and semiconductors.

**Figure Legends**

**Fig. 1. Focused helium ion beam Josephson junction fabrication.** (**a**) Photograph of a photolithographically patterned Au/YBCO film on a 5 mm × 5 mm sapphire substrate. (**b**) Zoomed view of the central substrate region showing where the gold contact region and YBCO film were etched prior to junction fabrication. This particular pattern contained four, 4 μm wide contacts for four-point resistance measurements. The helium ion beam was scanned along the red lines. The current ($I$) and voltage ($V$) leads for one of these bridge is labeled. (**c**) An artistic rendition of the focused helium ion beam creating a Josephson junction in the YBCO film. The crystal structure is that of YBCO.

**Fig. 2. Electrical transport measurements for a SNS Josephson junction fabricated using a dose of 2 × 10$^{16}$ He$^+$/cm$^2$.** (**a**) Current-Voltage characteristics measured for temperatures 63, 65, 67, 69, 71, and 75 K. (**b**) The Temperature dependence of the $I_C$ and $R_N$. (**c**) The Fraunhofer diffraction pattern for the critical current in magnetic field at 77 K. **Electrical transport measurements for a SIS Josephson junction fabricated using a dose of 6 × 10$^{16}$ He$^+$/cm$^2$.** (**d**) Current-Voltage characteristics measured at 6, 10, 12, 14, 16, 18, and 22 K. (**e**) The temperature dependence of $I_C$ and $R_N$. (**f**) The Fraunhofer diffraction pattern for the critical current in magnetic field at 6 K. (**g**) Current-Voltage characteristics for a high voltage range. (**h**) d$I$/d$V$ for the temperatures ranging from 70 to 6 K in 5 K increments. (**i**) Temperature dependence of 2Δ and BCS fit. (**j**) Zoomed in view of d$I$/d$V$ in the low bias regime around the super current.

**Methods**

YBCO films 150 nm thick were grown using reactive coevaporation on cerium oxide buffered sapphire wafers. For electrical contacts, 200 nm of gold was sputtered onto the film before breaking vacuum. Wafers were diced into 5 mm ´ 5 mm substrates and coated with a layer of

Fuji OCG825 photoresist. Using a contact mask aligner the base structure was exposed and developed using Fuji OCG934 developer. This pattern was transferred onto both the gold and underlying YBCO film using DC argon ion milling. A second lithography and ion milling step was performed to remove the gold contact layer and to reduce the thickness of the YBCO in the regions intended for junctions.

# Supplementary Information

**Extended Data Fig. 1** Simulated 30 keV He ion irradiation of a YBCO film using binary collision approximation (BCA) in Silvaco Athena. In each of these simulations we use 50,000,000 ion trajectories to simulate fluences of **(a)** $1\times10^{16}$, **(b)** $2\times10^{16}$, **(c)** $6\times10^{16}$, and **(d)** $20\times10^{16}$ ions/cm$^2$.

**Extended Data Fig. 2** Measurements for a YBCO sample irradiated with a relatively small dose of $1 \times 10^{16}$ He$^+$/cm$^2$. (a) $I$-$V$ characteristic with large excess current. (b) Resistance as a function of temperature.

**Extended Data Fig. 3** Measurements for a YBCO sample irradiated with a relatively high dose of $6 \times 10^{17}$ He$^+$/cm$^2$. **(a)** $I$-$V$ characteristic. **(b)** Resistance as a function of temperature The discontinuity near 85 K is the superconducting transition of the electrodes. .

**Extended Data Fig. 4** $I$-$V$ characteristic and $dV/dI$. measured at 6K showing Fiske mode cavity resonances with periodicity of 35 µV.

**Extended Data Fig. 5** High current measurements for a SIS Josephson junction. **(a)** $I$-$V$ characteristic showing the critical current of the electrodes and heating effects. **(b)** Critical current as a function of temperature.

**Acknowledgements:**
This work was supported by the Office of Science and Office of Basic Energy Sciences of the U.S. Department of Energy under Contract No. DEAC02 05CH11231 and by AFOSR grant FA9550-07-1-0493. M.M. and B.W. were supported by the UC scholars program. The authors gratefully acknowledge Garrett Schlenvogt for help with ion implantation simulations, Ke Chen and Peter Roediger for experimental discussions, Jhih-Sheng Wu for help with the BCS fit and K. D. Derr, Bernhard Goetze and John Notte, for helping with setting up this experiment.

**Author Contributions**
S.C. conceived the experiment and wrote the manuscript. S.C., E.C., and C.H. fabricated the devices. All of the authors provided technical and scientific insight that contributed to characterization of the devices and interpretation of the results.

**Author Information** The authors declare no competing financial interests.

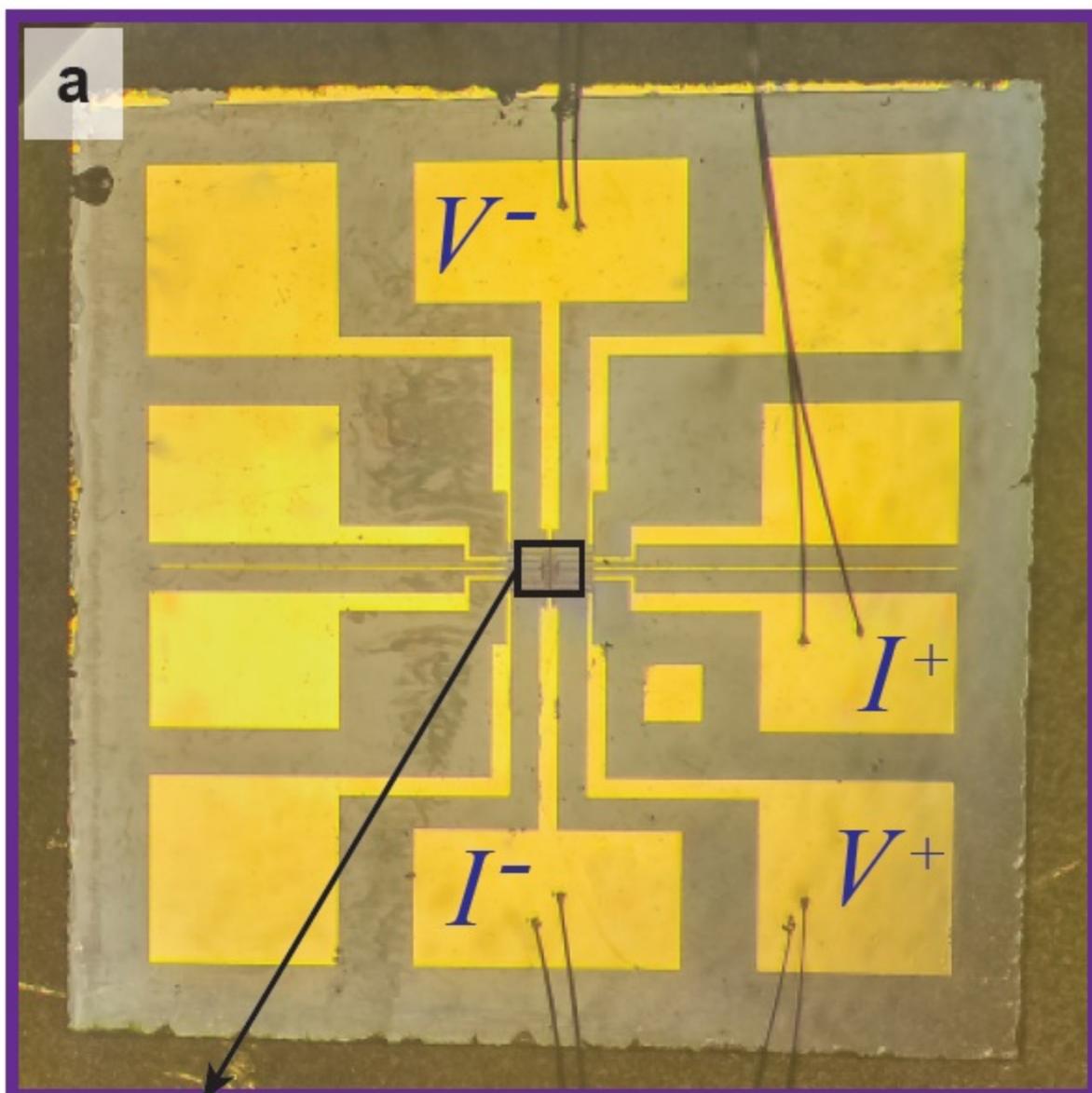
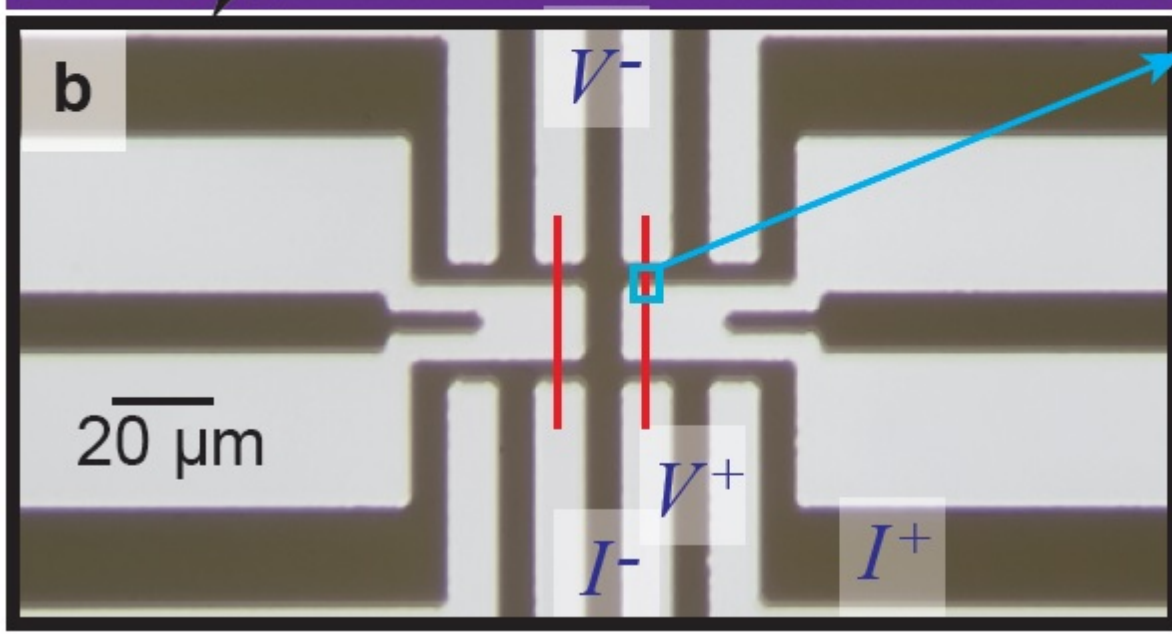
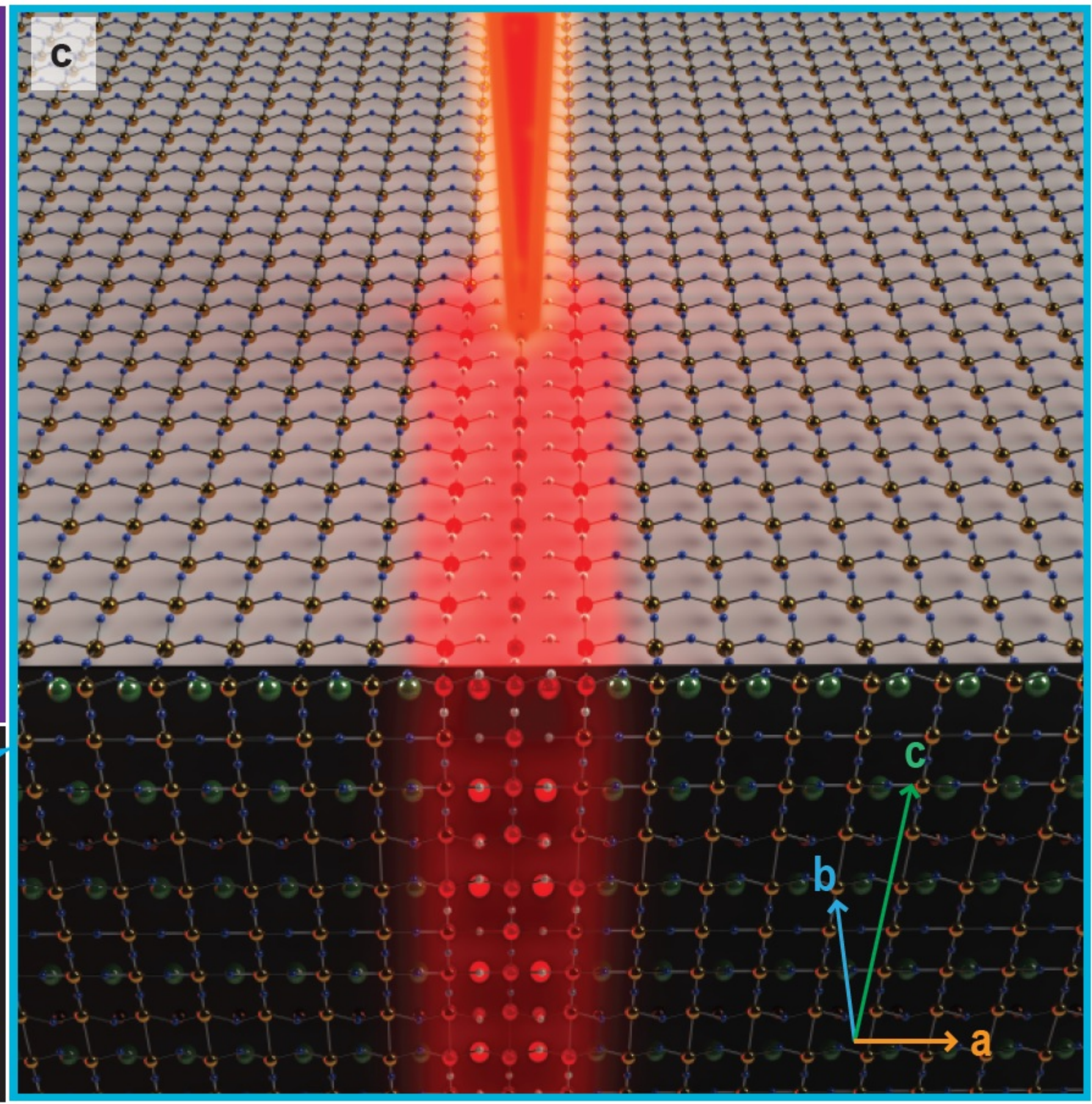

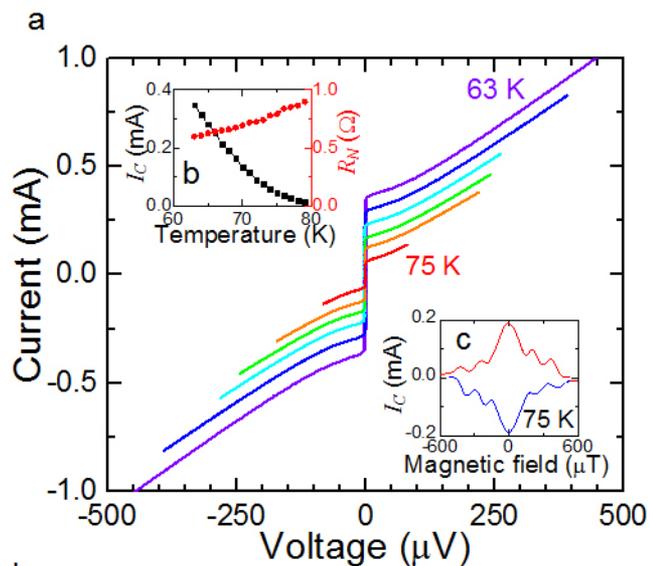
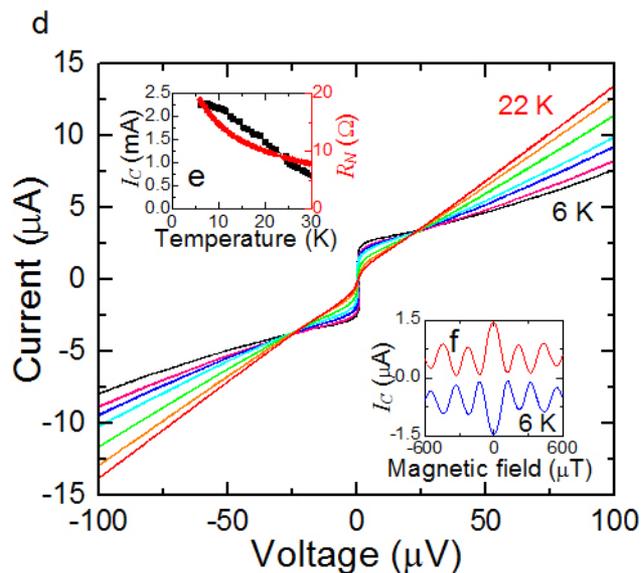
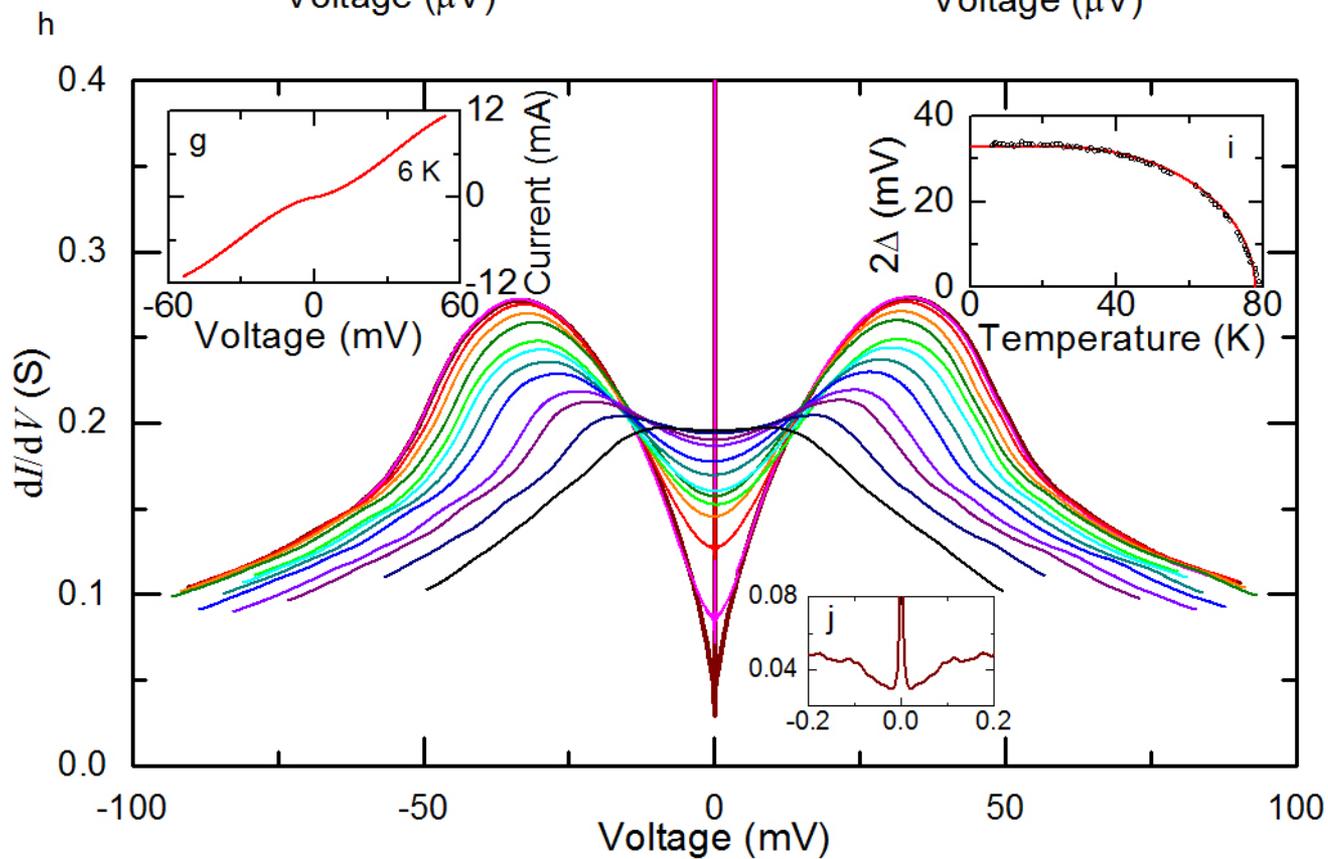

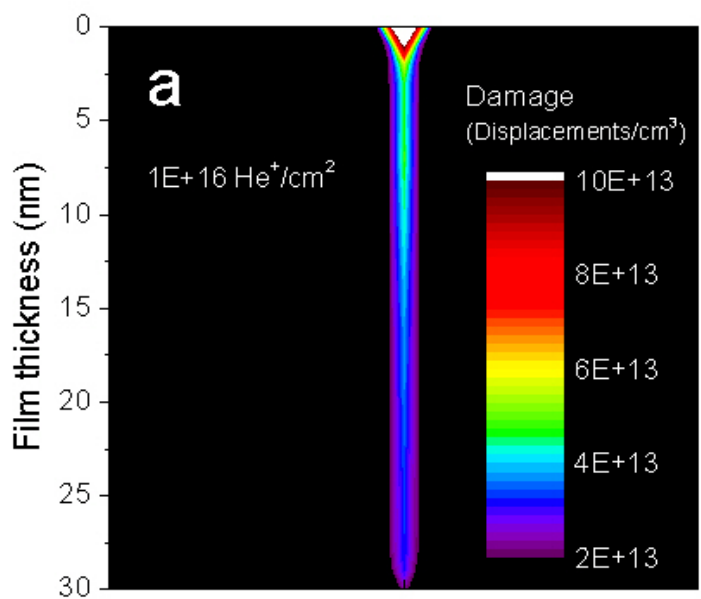
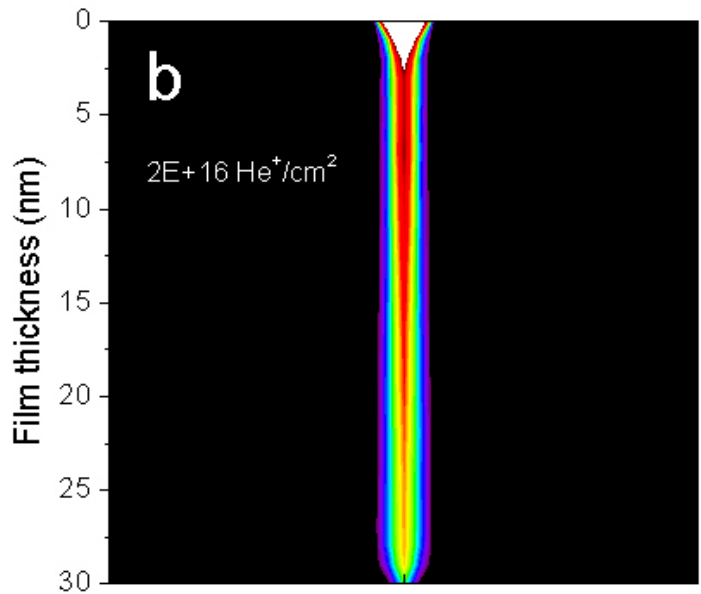
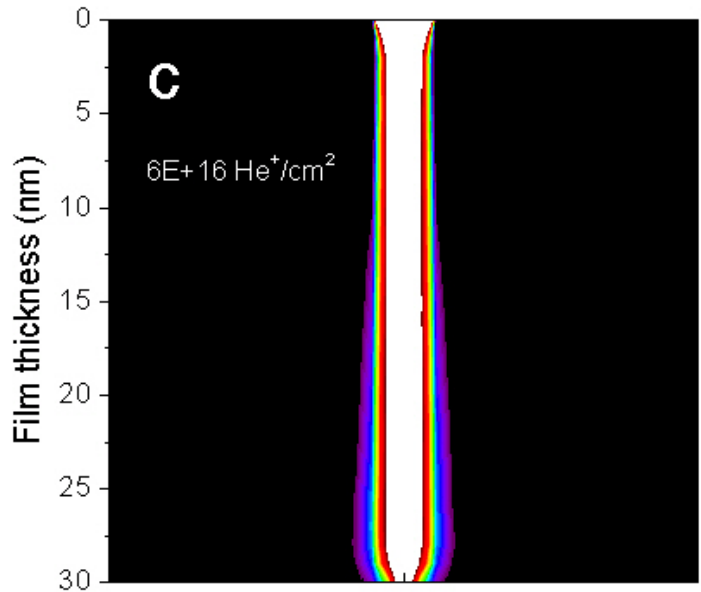
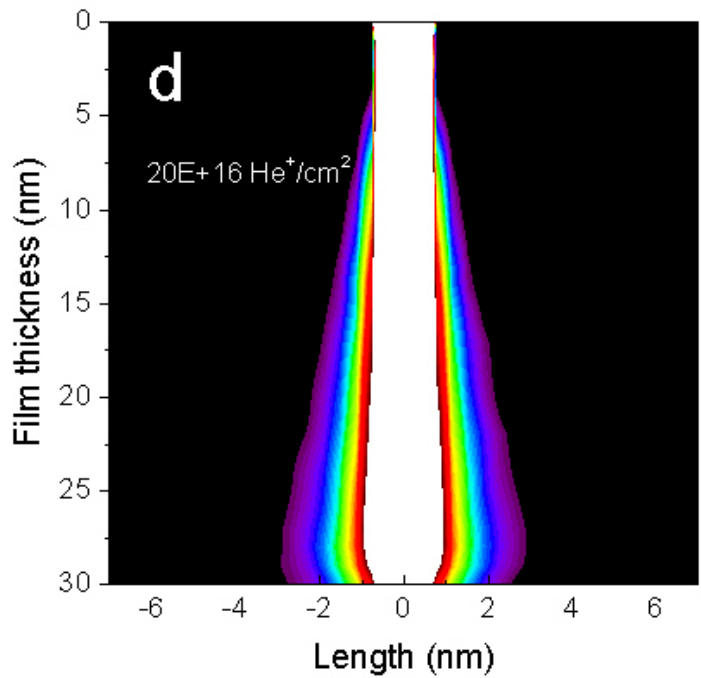

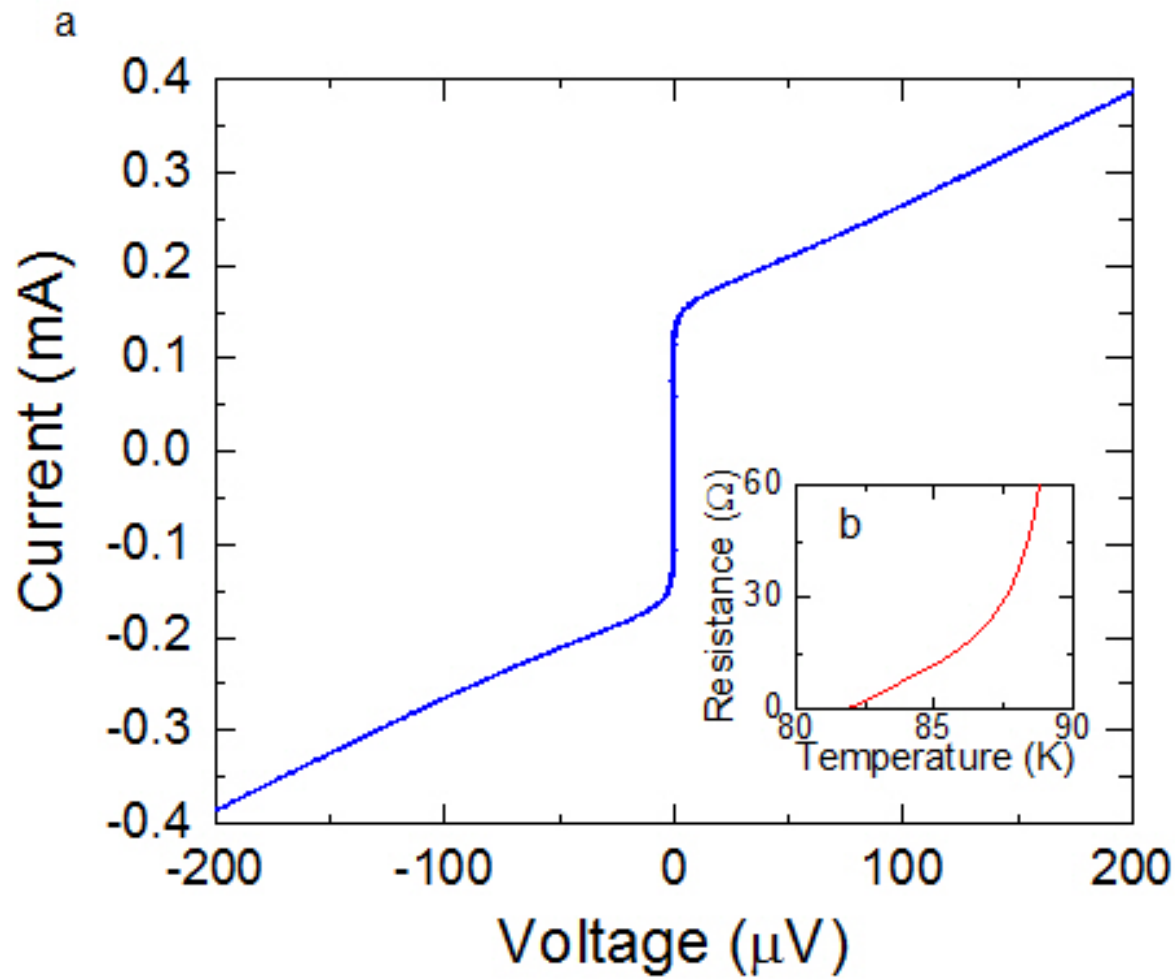

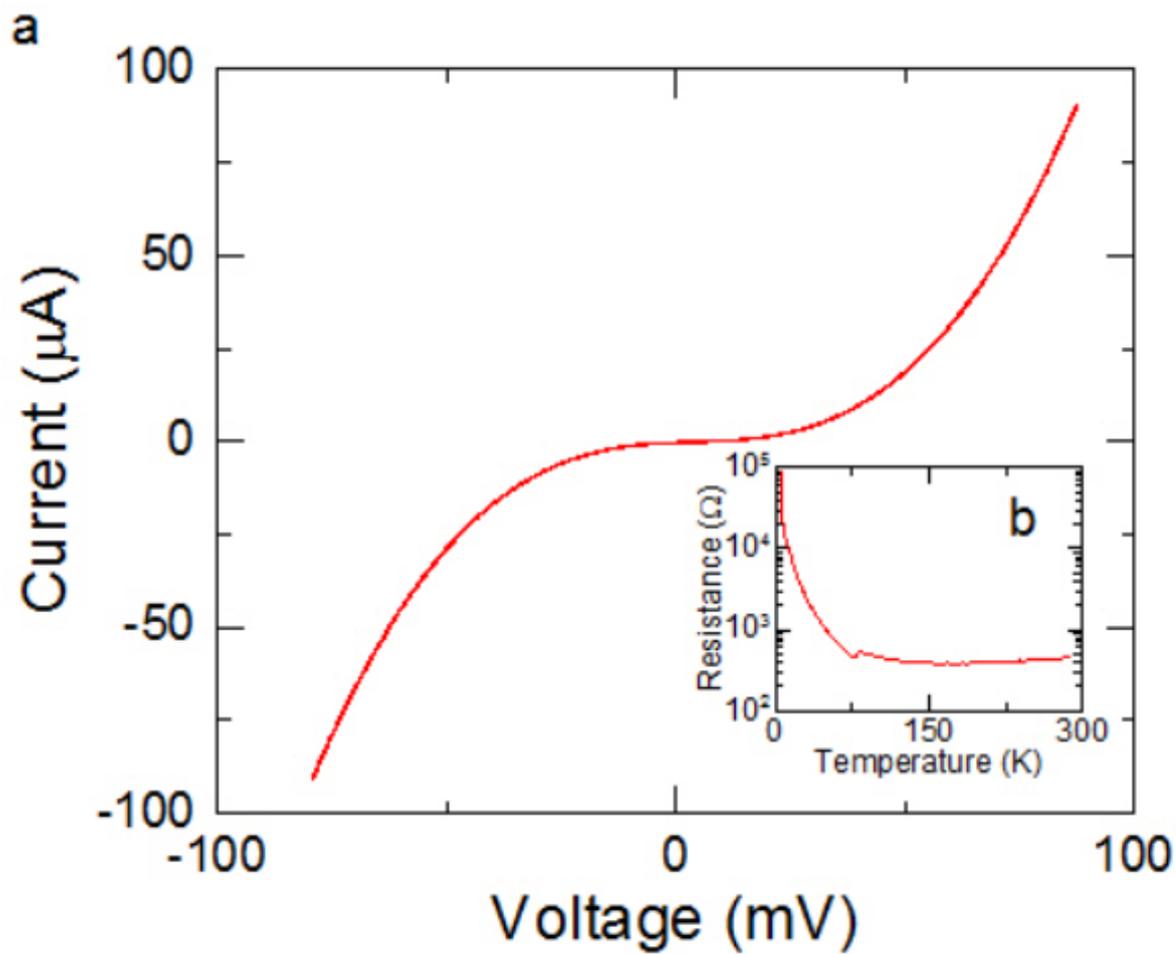

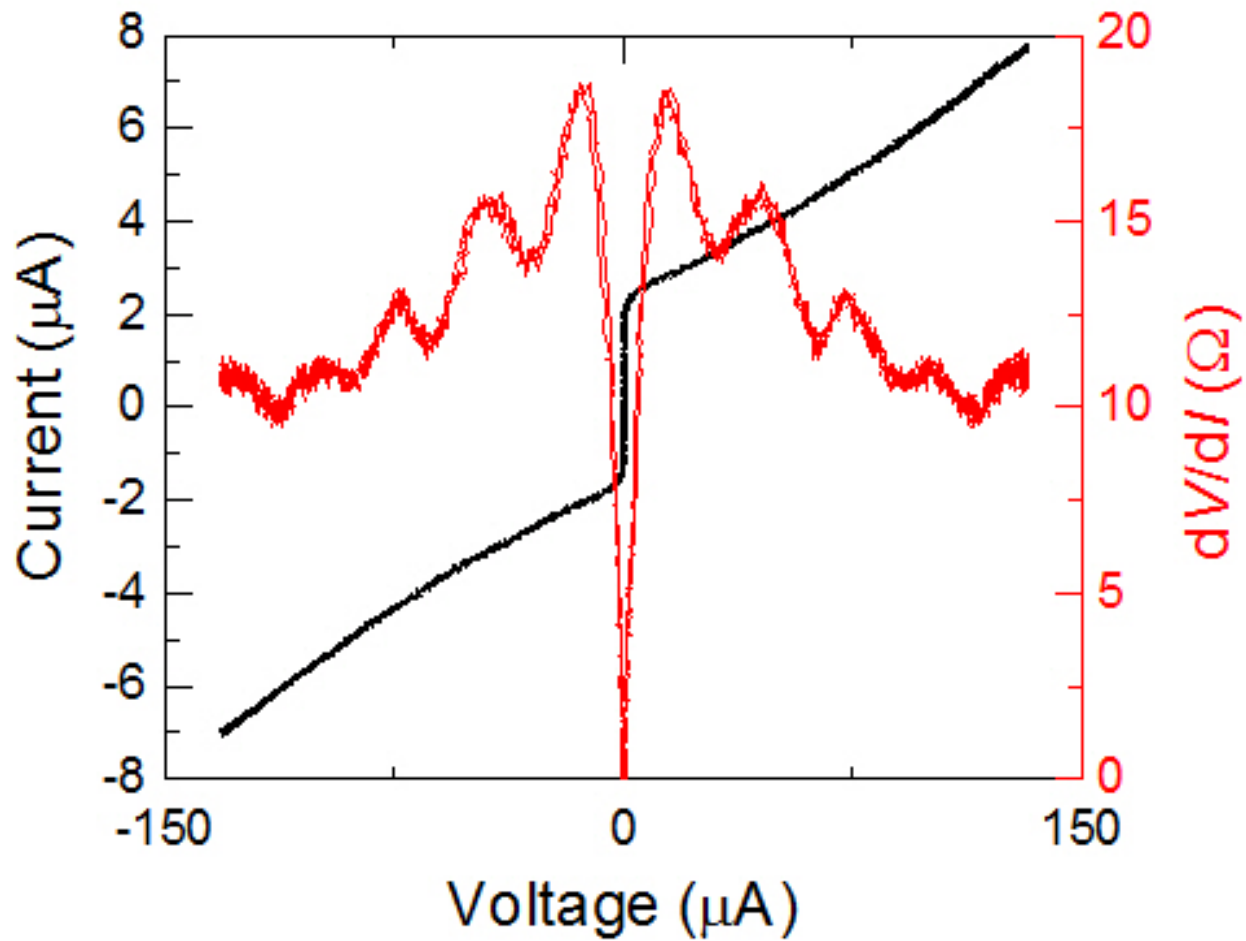

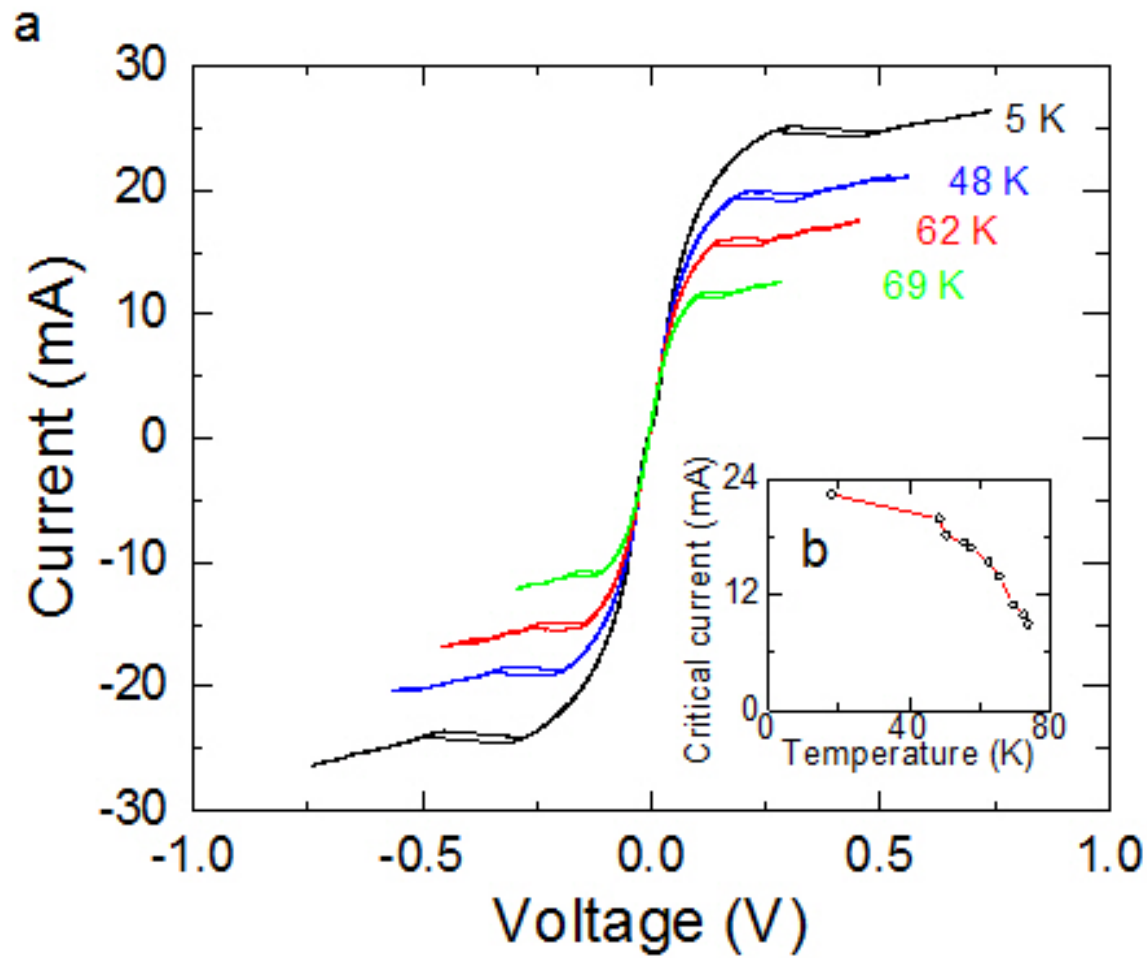